\pdfoutput=1
\documentclass{article}

    \usepackage[preprint]{neurips_2024}

\usepackage[utf8]{inputenc} 
\usepackage[T1]{fontenc}    
\usepackage{hyperref}       
\usepackage{url}            
\usepackage{booktabs}       
\usepackage{amsfonts}       
\usepackage{nicefrac}       
\usepackage{microtype}      
\usepackage{xcolor}         

\usepackage{amsmath}
\usepackage[normalem]{ulem}
\useunder{\uline}{\ul}{}
\usepackage{multirow}
\usepackage{xcolor}
\usepackage{subcaption}
\usepackage{graphicx}

\title{NumLLM: Numeric-Sensitive Large Language Model for Chinese Finance}

\author{%
	Huan-Yi Su, Ke Wu, Yu-Hao Huang, Wu-Jun Li$^*$\\
	National Key Laboratory for Novel Software Technology\\
	Department of Computer Science and Technology\\
	Nanjing University, Nanjing 210023, China \\
	\texttt{\{shyringo, ke.wu, huangyh\}@smail.nju.edu.cn,} \texttt{liwujun@nju.edu.cn}\\
}

\begin{document}

\maketitle
\def\thefootnote{*}\footnotetext{Corresponding author.}\def\thefootnote{\arabic{footnote}}

\begin{abstract}
  Recently, many works have proposed various financial large language models~(FinLLMs) by pre-training from scratch or fine-tuning open-sourced LLMs on financial corpora.
However, existing \mbox{FinLLMs} exhibit unsatisfactory performance in understanding financial text when numeric variables are involved in questions. In this paper, we propose a novel LLM, called \underline{num}eric-sensitive \underline{l}arge \underline{l}anguage \underline{m}odel~(NumLLM), for Chinese finance.
We first construct a financial corpus from financial textbooks which is essential for improving numeric capability of LLMs during fine-tuning. After that, we train two individual low-rank adaptation~(LoRA) modules by fine-tuning on our constructed financial corpus. One module is for adapting general-purpose LLMs to financial domain, and the other module is for enhancing the ability of NumLLM to understand financial text with numeric variables. Lastly, we merge the two LoRA modules into the foundation model to obtain NumLLM for inference. 
Experiments on financial question-answering benchmark show that \mbox{NumLLM}  can boost the performance of the foundation model and can achieve the best overall performance compared to all baselines, on both numeric and non-numeric questions.

\end{abstract}

\section{Introduction}

\noindent
Large language models~(LLMs), often comprising more than billions of parameters, have revolutionized the research paradigm in natural language processing~(NLP). 
By pre-training on massive corpora, LLMs have shown their excellent capability in learning complex language patterns and representations due to their immense model size~\citep{GPT3,LLaMA,LLAMA2}. 
LLMs have also shown promising performance in natural language understanding and generation tasks, such as question answering, machine translation and sentiment analysis~\citep{Baichuan2,LLaMA}. 
Hence, LLMs have attracted much attention in the artificial intelligence community.

Recently, many works have proposed various financial large language models~(FinLLMs) by pre-training from scratch or fine-tuning open-sourced LLMs on financial corpora. For example, BloombergGPT~\citep{BloombergGPT} and XuanYuan 2.0~\citep{XuanYuan} are pre-trained with a BLOOM-style~\citep{BLOOM} LLM from scratch. \mbox{DISC-FinLLM}~\citep{DISC-FinLLM}, FinMA~\citep{PIXIU}, Fin-Alpaca-LoRA-Linly~\citep{Cornucopia} and \mbox{FinGPT-v3}~\citep{FinGPT} are fine-tuned from Baichuan~\citep{Baichuan2}, LLaMA~\citep{LLaMA}, Chinese-LLaMA~\citep{Chinese-LLaMA} and \mbox{ChatGLM2}~\citep{GLM}, respectively. All these FinLLMs, except for \mbox{FinGPT-v3}, are pre-trained or fine-tuned on financial corpora collected by their corresponding authors.

Although these existing FinLLMs can achieve impressive performance in financial natural language understanding, they exhibit unsatisfactory performance in understanding financial text when numeric variables are involved in questions. More specifically, most of them, except for \mbox{FinGPT-v3}, are trained with next-token prediction objectives in an auto-regressive manner, which only includes preceding context for prediction of numeric variables. However, training in an auto-regressive manner cannot fully learn the context dependency of numeric variables~\citep{GLM} which is important for understanding financial text with numeric variables. Although FinGPT-v3 can learn the context dependency with an auto-regressive blank infilling objective, it constructs blank tokens with random masking, lacking sensitivity to numeric variables within financial text. Since it is common for financial text to involve numeric variables, improving the numeric capability is essential for \mbox{FinLLMs} to better understand financial text with numeric variables.

In this paper, we propose a novel LLM, called \underline{num}eric-sensitive \underline{l}arge \underline{l}anguage \underline{m}odel~(NumLLM), for Chinese finance~\footnote{We focus on Chinese finance in this paper. But the techniques proposed in this paper can be easily adapted to finance in other languages.}. The contributions of this paper are outlined as follows:
\begin{itemize}
    \item We construct a financial corpus from financial textbooks, which is essential for improving numeric capability of LLMs during fine-tuning.
    \item We develop a novel fine-tuning method with two individual low-rank adaptation~(LoRA) modules to enhance the ability of NumLLM in understanding financial text with numeric variables.

    \item Experiments on financial question-answering benchmark show that \mbox{NumLLM}  can boost the performance of the foundation model and can achieve the best overall performance compared to all baselines, on both numeric and non-numeric questions.
\end{itemize}

\section{Related Works}

In this section, we introduce some related works about financial corpora and financial LLMs.
 
\subsection{Financial Corpora}

Adapting LLMs for a particular domain often requires domain-specific corpora. Therefore, constructing financial corpora is a crucial step for training financial LLMs. Existing works have constructed a few financial corpora in various ways. For example, FinGPT-v3~\citep{FinGPT} constructs its financial corpora from diverse sources, such as financial news, filing data and social media, which can be collected from Stocknet~\citep{Stocknet} and FiQA SA~\citep{FiQA-SA}. BBT-FinCorpus~\citep{BBT-Fin} is a massive Chinese financial corpus, collected from financial news, company announcements, research reports, and social media. TigerBot~\citep{TigerBot} constructs its corpus from thousands of research reports and earnings reports. Yayi~\footnote{\url{https://huggingface.co/datasets/wenge-research/yayi_domain_subset}} is an instruction tuning dataset which is constructed from financial news events. DISC-Fin-SFT~\citep{DISC-FinLLM} is an instruction dataset derived from various data sources. PIXIU~\citep{PIXIU} constructs a financial instruction tuning dataset~(FIT) from open-sourced data. 

All financial corpora mentioned above lack financial expertise from financial textbooks. This phenomenon motivates us to construct a financial corpus collected from financial textbooks which is essential for improving numeric capability of LLMs during fine-tuning.

\subsection{Financial LLMs}
Since general-purpose LLMs are pre-trained on massive and diverse corpora to learn general language representations, fine-tuning is often required to adapt them to specific domains. Existing financial LLMs can be mainly categorized into models pre-trained from scratch and models fine-tuned from open-sourced LLMs. Models pre-trained from scratch include BloombergGPT~\citep{BloombergGPT} and XuanYuan 2.0~\citep{XuanYuan}, both of which are BLOOM-style~\citep{BLOOM} LLMs. More specifically, BloomberGPT pre-trains a BLOOM-50B model on its collected massive financial corpora. XuanYuan 2.0 pre-trains a BLOOM-176B model on its collected Chinese financial corpora. Models fine-tuned from open-sourced LLMs include \mbox{DISC-FinLLM}~\citep{DISC-FinLLM}, PIXIU~\citep{PIXIU}, Fin-Alpaca-LoRA-Linly~\citep{Cornucopia} and \mbox{FinGPT-v3}~\citep{FinGPT}, which are fine-tuned from different open-sourced LLMs. For example, \mbox{DISC-FinLLM} is fine-tuned from Baichuan 13B~\citep{Baichuan2} with its proposed multiple experts fine-tuning framework. PIXIU, the first English financial LLM, is fine-tuned from LLaMA~\citep{LLaMA} with its constructed instruction data. Fin-Alpaca-LoRA-Linly, a model for question-answering in Chinese finance, is fine-tuned from Chinese-LLaMA~\citep{Chinese-LLaMA} which is a LLaMA model adapted for Chinese. FinGPT-v3 applies LoRA~\citep{LoRA} to fine-tune \mbox{ChatGLM2}~\citep{GLM} with the inherent feedback from markets. XuanYuan 2.0, DISC-FinLLM, Fin-Alpaca-LoRA-Linly and FinGPT-v3 are for Chinese finance, while BloomberGPT and PIXIU are for finance tasks in other languages.

Although existing financial LLMs can achieve good performance in financial natural language understanding tasks, they exhibit unsatisfactory performance in understanding financial text when numeric variables are involved in questions. This phenomenon motivates the work in this paper.

\begin{figure*}[t]
    \centering
    \includegraphics[width=\textwidth]{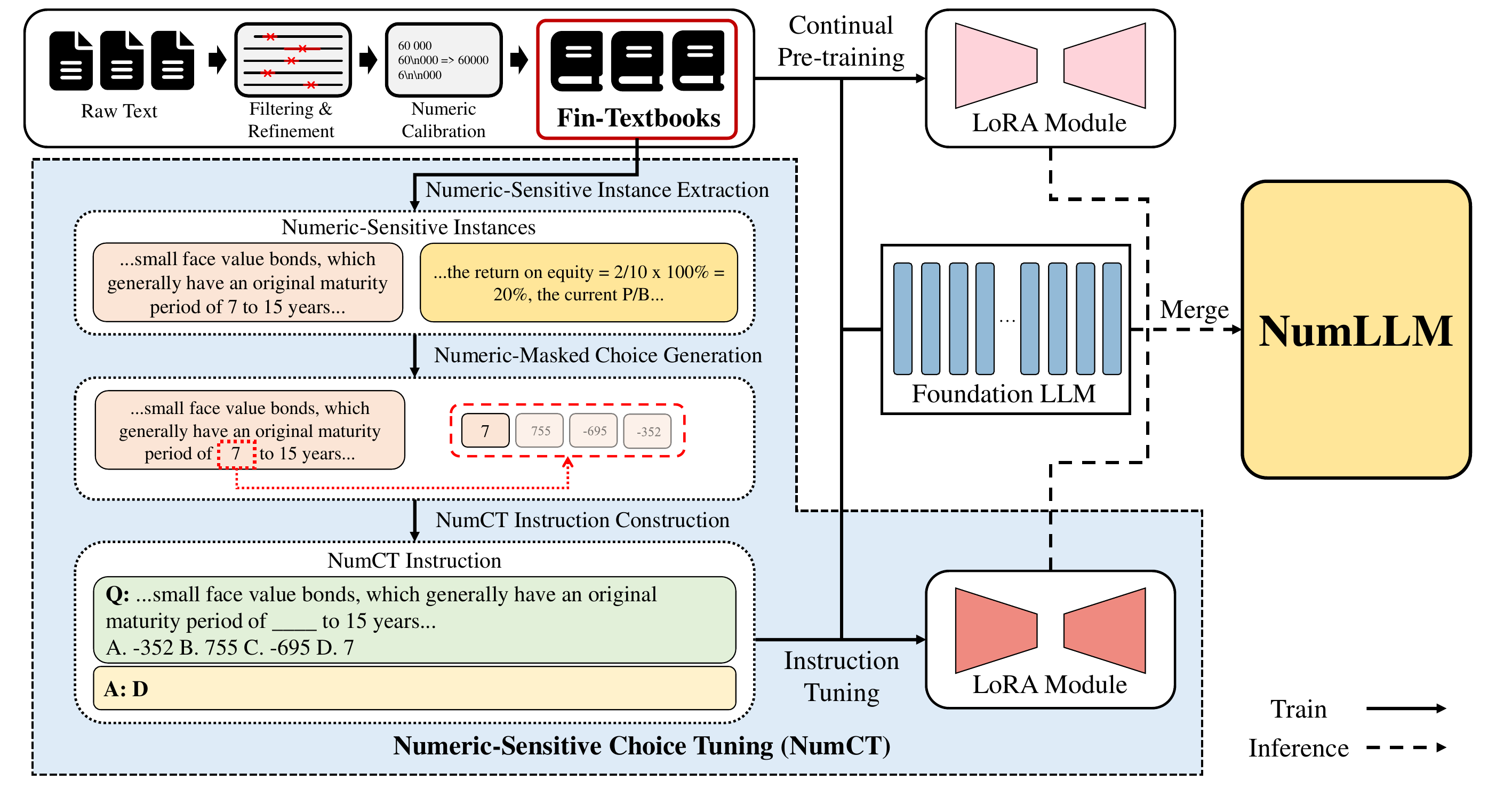}
    \caption{The architecture of NumLLM.}
    \label{figure-方法总览图}
\end{figure*}

\section{Numeric-Sensitive Large Language Model}

In this section, we introduce the details of our proposed \mbox{NumLLM}, the architecture of which is illustrated in Figure~\ref{figure-方法总览图}. Firstly, we construct a financial corpus, called Fin-Textbooks, from textbooks in finance. After that, we train two individual LoRA modules by fine-tuning on \mbox{Fin-Textbooks}. In particular, one module is for continual pre-training by fine-tuning the foundation LLM with next-token prediction task. 
The other module is trained by fine-tuning the foundation model with our proposed \underline{num}eric-sensitive \underline{c}hoice \underline{t}uning~(NumCT) to enhance the capability of the LLM in understanding financial text with numeric variables. 
 Lastly, we mix the two LoRA modules and merge the mixed LoRA module into the foundation model to obtain \mbox{NumLLM} for inference. We choose Qwen-7B~\citep{Qwen} as the foundation model, because our experiments show that Qwen-7B is superior over other models with comparable model size on both numeric and non-numeric questions.

\subsection{Fin-Textbooks: Chinese Financial Textbook Corpus}
\label{section:Fin-Textbooks}
Fin-Textbooks consists of 24 preprocessed financial textbook documents. It covers 34 different financial subjects, including fundamentals of futures and derivatives, probability and mathematical statistics and so on. The statistics of Fin-Textbooks are summarized in Table~\ref{summary of Fin-Textbooks}. All textbooks are crawled or downloaded from websites.

\begin{table}[!t]
\small
\centering 
\caption{Statistics of Fin-Textbooks}
\begin{tabular}{lccc}
\hline
{Item} & {Value}  \\ \hline
Number of subjects & 34  \\
Number of documents & 24  \\
Number of tokens & 6,913,132 \\ \hline
\end{tabular}

\label{summary of Fin-Textbooks}
\end{table}

We preprocess the raw textbooks by filtering, refinement and numeric calibration. The details are as follows:
\begin{itemize}
    \item The filtering operation removes non-financial content from the raw textbooks, such as  information of publication and list of references. 
    \item The refinement operation further eliminates components that do not contain financial knowledge, such as table of contents and some section headings.
    \item The numeric calibration addresses numeric-related formatting issues in the raw textbook texts, such as spacing and paragraph breaks within numeric variables.
\end{itemize}

\subsection{Continual Pre-Training} 
Continual pre-training refers to domain-adaptive pre-training with augmented data~\citep{DBLP:conf/acl/GururanganMSLBD20}. Continual pre-training has been proved successful in adapting pre-trained language models to domain-specific tasks~\citep{DBLP:conf/aaai/0008GRPRDGY23,DBLP:conf/kdd/XieLLXWN23,DBLP:conf/acl/GongZZSWW22}. We apply LoRA to continually pre-train Qwen-7B on Fin-Textbooks. The training settings are the same as in Qwen-7B. The learning task is to perform next-token prediction as in the standard language modeling objective~\citep{palm}. In particular, we maximize the following log likelihood function:
 \begin{equation}
 L_{\text{CP}}=\sum_i\log P(w_i|w_{i-k},\ldots,w_{i-1};\Theta),
 \end{equation}
 where $w_i$ is the $i$-th token in the corpus, $k$ is the size of the context window and $\Theta$ is the model parameters.

\subsection{Numeric-Sensitive Choice Tuning}

NumCT is developed to enhance the capability of the LLM in understanding financial text when numeric variables are involved in questions. NumCT includes four steps: numeric-sensitive instance extraction, numeric-masked choice generation, NumCT instruction construction and instruction fine-tuning.

\subsubsection{Numeric-Sensitive Instance Extraction}
\label{section:r_ins}
In this step, we extract instances containing numeric variables from the preprocessed corpus, where each instance is a segment of text. We define the hyperparameter $n_{min}$ as the minimum number of paragraphs per instance and $n_{max}$ as the maximum number of paragraphs per instance. These two  hyperparameters influence the average length per instruction. We define $r_{\text{ins}}$ as the ratio of selected instances. We conduct instance extraction from the beginning of the corpus. For each instance, we initialize it as an empty string and add $n_{min}$ paragraphs in the first place. We make sure that each instance is grammatically intact and does not exceed $n_{max}$ paragraphs. If an instance does not contain numeric variables, it is discarded. In addition, if all the numeric variables are structural variables, such as the ``3'' in ``Figure 3'', the instance is discarded. We repeat this procedure until we reach the end of the corpus.
 
After going through the whole corpus, we can extract $N_{\text{ins}}$ instances. In the end, we randomly select $r_{\text{ins}}$ portion of the instances, which is
$\tilde{N}_{\text{ins}}=\lceil r_{\text{ins}}\times N_{\text{ins}}\rceil$
instances, for the next step. The randomness in numeric-sensitive instance extraction enhances the relevance of financial knowledge in the selected instances. Even in textbooks, there are still rare texts that are irrelevant and contain little financial knowledge, like common formal expressions and nonessential details in the used examples. These irrelevant texts are not likely to be removed in the preprocessing stage because of the variety of structures and styles across different textbooks. If we assume the number of instances composed of such irrelevant texts is $N_\text{irr}\ll N_{\text{ins}}$, the probability of all the selected instances being relevant should be 
 \begin{equation}p
 =\frac
 {{ N_{\text{ins}}-N_\text{irr} \choose \tilde{N}_{\text{ins}} }}
 { {{N_{\text{ins}} \choose \tilde{N}_{\text{ins}}}}}
 =
 \frac
 {(N_{\text{ins}}-N_\text{irr})!}
 {N_{\text{ins}}!}
 \prod^{N_\text{irr}}_{j=1}(N_{\text{ins}}-N_\text{irr}-\tilde{N}_{\text{ins}}+j).
  \end{equation}
 Since $\tilde{N}_{\text{ins}}=\lceil r_{\text{ins}} \times N_{\text{ins}}\rceil $, a smaller $r_{\text{ins}}$ means a larger $p$ and lower occurrence of irrelevant texts. Please note that $r_{\text{ins}}$ should not be too small, in order to make full usage of the corpus.

 \subsubsection{Numeric-Masked Choice Generation}
 \label{section:r_NV}
We define $r_{\text{NV}}$ as the ratio of selected numeric variables to mask per instance and $n_{\text{cho}}$ as the number of choices in each instruction. For each instance $I_t$, where $t=1,2,\dots,\tilde{N}_{\text{ins}}$, we perform numeric-masked choice generation. Suppose there are $M_t$ legitimate numeric variables in the current instance $I_t$. The numeric variables with the same numeric value but at different positions within the instance will be treated as different numeric variables. We then randomly select 
$\tilde{M}_t=\lceil r_{\text{NV}}\times M_t\rceil$
 numeric variables from $I_t$ for numeric-masked choice generation. For each numeric variable $\text{NV}_{ti}\in\bigcup\limits_{i=1}^{\tilde{M}_t}\{\text{NV}_{ti}\},$ we define $v_{ti}$ as its numeric value. For each $\text{NV}_{ti}$, we generate $(n_{\text{cho}}-1)$ numeric choices, denoted as $\{c_{tij}\}$, where $j=1,2,\dots,n_{\text{cho}}-1$. Specifically, we handle $\text{NV}_{ti}$ in two different ways according to its numeric type, thus enabling the LLM to learn to reason on both integers and floating-point numbers. If $v_{ti}$ is a floating-point number, we generate $(n_{\text{cho}}-1)$  random floating-point numbers within the following interval: 
 \begin{equation}
 c_{tij}\in\left[\lfloor v_{ti} \rfloor,\lfloor v_{ti}+1 \rfloor\right],j=1,2,\dots,n_{\text{cho}}-1.
  \end{equation}
 If $v_{ti}$ is an integer, we generate $(n_{\text{cho}}-1)$  random integers between the following interval:
 \begin{equation}
 c_{tij}\in\left[-s\left|v_{ti}\right|, s\left|v_{ti}\right|\right],j=1,2,\dots,n_{\text{cho}}-1,
  \end{equation}
 where $s>0$ is a scaler and set to be 1000 in our implementation. The randomness in numeric-masked choice generation maintains the diversity of instructions, which can improve model performance according to LIMA~\citep{LIMA}. The appropriate value of $r_{\text{NV}}$ is dependent on the corpus. If $r_{\text{NV}}$ is set to be too large, then most of the content of the instructions constructed from the same instance would be overlapped, thus impairing diversity. $r_{\text{NV}}$ should not be too small either, in order to make full exploitation of the corpus.
\subsubsection{NumCT Instruction Construction}

One NumCT instruction is a string comprised of a question, $n_{\text{cho}}$ identifiers $\{\text{ID}_k\}$, $n_{\text{cho}}$ choices $\{C_k\}$, and the necessary prompt constituents, where $k=1,2,\dots,n_{\text{cho}}$. $\{\text{ID}_k\}$ corresponds to $\{C_k\}$, respectively. Then, for each $\text{NV}_{ti}$, we randomly select one identifier $\text{ID}_{k_{ti}}$ as the identifier for the choice of the correct answer. The randomness in selection is the same as the choice shuffling proposed in \mbox{Medprompt}~\citep{Medprompt}, which can be helpful in mitigating position bias of models~\citep{DBLP:conf/emnlp/KoLKKK20,MT-Bench}. Then $C_{k_{ti}}$ is assigned as $v_{ti}$. The other choices, i.e., $\{C_k\},k=1,\dots,k_{ti}-1,k_{ti}+1,\dots,n_{\text{cho}}$, are assigned as $\{c_{tij}\},j=1,2,\dots,n_{\text{cho}}-1$, respectively. For $k=1,2,\dots,n_{\text{cho}}$, we concatenate $\text{ID}_k$ with $C_k$ to produce $F_k$.

\begin{figure}[t]
    \centering
    \includegraphics[width=0.5\linewidth]{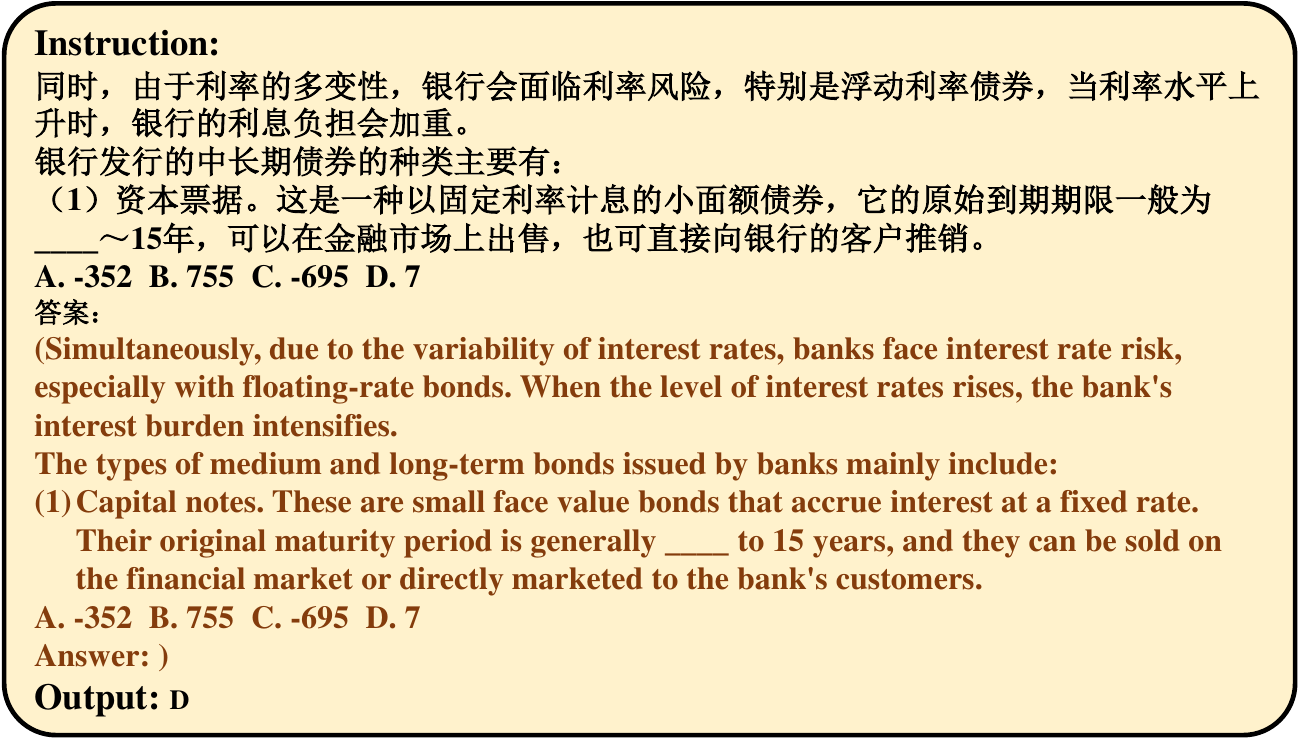}
    \caption{An example of instruction-output pair constructed by NumCT. Translation in English is provided below the original text.}
    \label{figure-NumCT生成的instruction-output的例子}
\end{figure}

Finally, for each $\text{NV}_{ti}$, we transform the instance into a question by masking $\text{NV}_{ti}$ with a blank underline of four token length. For $\text{NV}_{ti}$, we generate an NumCT instruction, by combining the question, $\{F_k\},k=1,2,\dots,n_{\text{cho}}$ and the necessary prompt constituents. The output matching the instruction is $\text{ID}_{k_{ti}}$. As a common practice~\citep{MMLU,C-Eval}, we set $n_{\text{cho}}=4$ and set the identifiers as ``A'', ``B'', ``C'', ``D''. An example of the instruction-output pair is shown in Figure \ref{figure-NumCT生成的instruction-output的例子}.

\subsubsection{Instruction Fine-Tuning}
After the above steps, we obtain an instruction-output pair for each $\text{NV}_{ti}$. By traversing all the selected numeric variables in all the selected instances, we obtain an instruction fine-tuning dataset containing $N$ instruction-output pairs, where $N$ is computed as:
 \begin{equation}
 N=\sum\limits^{\tilde{N}_{\text{ins}}}_{t=1}\tilde{M}_t.
  \end{equation}
We use this instruction fine-tuning dataset to perform instruction fine-tuning~\citep{FLAN} on the foundation LLM. The settings of fine-tuning are the same as the standard settings of fine-tuning Qwen, LLAMA2 and so on. We optimize an auto-regressive objective function, while zeroing out the loss on tokens from the instruction. NumCT maximizes the following log likelihood function:
\begin{equation}
L_{\text{NumCT}}=\sum_{j=1}^{N}\sum_{i=1}^{\tilde{l}_j}\log P(o_i), 
 \end{equation}
where
\begin{equation}
P(o_i)=
    \begin{cases}
    P\left(o_i|w_{l_j+i-k},\dots,w_{l_j},o_1,\dots,o_{i-1};\Theta\right)&,i\leq k \\
    P\left(o_i|o_{i-k},\dots,o_{i-1};\Theta\right)&,i>k
    \end{cases}.
 \end{equation}
Here, $N$ is the number of instruction-output pairs, $l_j$ is the length of the $j$-th instruction, $\tilde{l}_j$ is the length of the $j$-th output, $w_{l_j}$ is the ${l_j}$-th token in the instruction, $o_i$ is the $i$-th token in the output, $k$ is the size of the context window and $\Theta$ is the model parameters.

\subsection{Mixing and Merging LoRA Modules}
\label{section:mixing lora}
After continual pre-training and NumCT, we obtain two LoRA modules. In the mixing and merging step, we employ a singular value decomposition~(SVD) based method to mix the two LoRA modules and finally merge LoRA modules into the foundation LLM with an add operation as in PEFT~\citep{peft}. For convenience, we denote the LoRA module of continual pre-training by $M_{\text{CP}}$ and denote the LoRA module of NumCT by $M_{\text{NumCT}}$. $r_{\text{CP}}$ is the rank of $M_{\text{CP}}$ and $r_{\text{NumCT}}$ is the rank of $M_{\text{NumCT}}$. $\Delta W_{\text{CP}}$ denotes the product between the two low-rank matrices learned for $M_{\text{CP}}$, and $\Delta W_{\text{NumCT}}$ denotes the product between the two low-rank matrices learned for $M_{\text{NumCT}}$. Let $r=max(r_{\text{NumCT}},r_{\text{CP}})$. We perform SVD on 
\begin{equation}
\Delta W_{\text{mean}}=\frac{\Delta W_{\text{NumCT}}+\Delta W_{\text{CP}}}{2},
 \end{equation}
and retain the top $r$ singular values for the mixed LoRA module. Specifically, SVD decomposition for $\Delta W_{\text{mean}}\in\mathbb{R}^{d\times k}$ can be represented by 
\begin{equation}
\Delta W_{\text{mean}}=U\Sigma V^T, U\in\mathbb{R}^{d\times d}, \Sigma\in\mathbb{R}^{d\times k}, V\in\mathbb{R}^{k\times k}.
 \end{equation} 
After extracting the top $r$ singular values and the corresponding singular vectors, we can obtain $U'\in\mathbb{R}^{d\times r}, \Sigma'\in\mathbb{R}^{r\times r}, V'\in\mathbb{R}^{k\times r}$. The we can obtain the full matrix for the mixed LoRA module as follows:
\begin{equation}
\Delta W_{\text{SVD}}=U'\Sigma' (V')^T.
 \end{equation} 
During inference, we merge the mixed LoRA module with the foundation model by using an add operation to obtain the NumLLM model, which is consistent with the default operation in LoRA~\citep{LoRA}. We set $r_{\text{CP}}=64$ and $r_{\text{NumCT}}=8$. By mixing and merging the two LoRA modules through an SVD-based method, we preserve the most important information from each LoRA module. Thus we enhance the ability of NumLLM to understand the financial texts involving numeric variables as well as those not involving numeric variables.

\section{Experiment}
In this section, we conduct experiments to compare our \mbox{NumLLM} with existing LLMs, including representative general-purpose LLMs and financial LLMs.
\subsection{Experimental Setup}

\subsubsection{Evaluation Tasks}
\label{section:Evaluation Tasks}
\begin{figure*}[t]
    \centering
    \includegraphics[width=\textwidth]{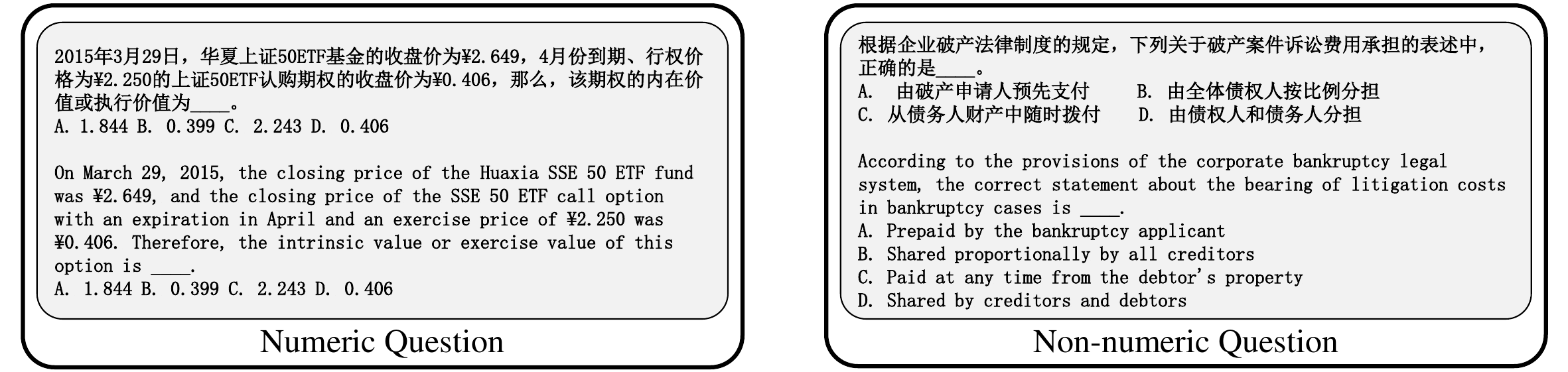}
    \caption{Examples of numeric and non-numeric questions. Translation in English is provided below the original text.}
    \label{figure-对比numeric和non-numeric问题}
\end{figure*}

We evaluate all models on FinEval~\citep{FinEval} which is a comprehensive benchmark for the Chinese financial question answering task. Each task is in the form of multiple-choice question answering and is evaluated under a five-shot scenario without chain-of-thought. Each task adopts accuracy as the evaluation metric.
As stated in FinEval, such an evaluation setting is reasonable because all methods can achieve the highest accuracy compared to that under the setting of zero-shot or chain-of-thought.
We present results in four sub-domains of finance, including Finance, Economy, Accounting and Certificate. We also present average results over all sub-domains.
Additionally, we decompose all questions within each sub-domain into numeric questions and non-numeric questions, and present the results respectively. Figure~\ref{figure-对比numeric和non-numeric问题} shows examples of numeric and non-numeric questions. More examples can be found in the Appendix. 
FinEval adopts the same settings as in existing works~\citep{MMLU,GPT3}, where the choice corresponding to the largest output logit is returned as the choice made by LLMs.
The prompt template in FinEval simply concatenates the necessary prompt constituents, the question, four identifiers and four choices. The question is masked partially with a blank underline of four token length.

The financial-domain questions in FinEval include 34 distinct subjects which are classified into four sub-domains. Please note that the testing set we use corresponds to the validation set in the original paper of FinEval, because the labels for the testing set in the original paper are not publicly available. The number of questions within the testing set is shown in Table \ref{statistics of FinEval}.

\begin{table}[!t]
\small
\centering 

\caption{Number of questions in FinEval. Each column denotes a sub-domain. ``All'' denotes the number of questions within all sub-domains. ``\#Numeric'' denotes the number of numeric questions and ``\#Non-numeric'' denotes the number of non-numeric questions. ``\#Total'' denotes the sum of ``\#Numeric'' and ``\#Non-numeric''.
}
\begin{tabular}{@{}lccccc@{}}
\toprule
            & Accounting & Certificate & Economy & Finance & All \\ \midrule
\#Numeric     & 68         & 73          & 42      & 56      & 239   \\
\#Non-numeric & 237        & 261         & 165     & 249     & 912   \\
\#Total       & 305        & 334         & 207     & 305     & 1151  \\ \bottomrule
\end{tabular}

\label{statistics of FinEval}
\end{table}

\begin{table*}[t]
\small
\centering 
\caption{Accuracy (\%) in four sub-domains and on FinEval overall. * indicates that the result of the model is directly adopted from the paper of FinEval, with one digit after the decimal point. In the column ``category'' , ``g'' means general-purpose LLM, ``f'' means financial LLM. The ``Overall'' accuracy is the average accuracy over all the subjects regardless of sub-domain, computed in the same way as in FinEval. The column ``n'' represents numeric questions, ``non-n'' represents non-numeric questions, and ``avg'' represents average accuracy over all the questions, whether numeric or non-numeric questions. Bold indicates the best result. Underline indicates the second best result. The numbers within parentheses are the standard deviations of NumLLM.}

\resizebox{\linewidth}{!}{
\begin{tabular}{@{}lcc|ccc|ccc|ccc|ccc|ccc@{}}
\toprule
\multirow{2}{*}{model}      & \multirow{2}{*}{size} & \multirow{2}{*}{category} & \multicolumn{3}{c|}{Accounting}                   & \multicolumn{3}{c|}{Certificate}                  & \multicolumn{3}{c|}{Economy}                      & \multicolumn{3}{c|}{Finance}                      & \multicolumn{3}{c}{Overall}                      \\
                            &                       &                           & n              & non-n          & avg            & n              & non-n          & avg            & n              & non-n          & avg            & n              & non-n          & avg            & n              & non-n          & avg            \\ \midrule
ChatGLM2                    & 6B                    & g                         & 35.29          & 59.92          & 54.43          & 32.88          & 58.24          & 52.69          & 35.71          & 44.85          & 43.00          & {\ul 39.29}    & 57.83          & 54.43          & 35.56          & 55.37          & 51.87          \\
ChatGLM3                    & 6B                    & g                         & 36.76          & 47.26          & 44.92          & 32.88          & 54.41          & 49.70          & 30.95          & 51.52          & 47.34          & 37.50          & 53.41          & 50.49          & 34.73          & 52.47          & 48.22          \\
LLaMA                       & 7B                    & g                         & \textbf{45.59} & 31.22          & 34.43          & 27.40          & 24.52          & 25.15          & 28.57          & 26.67          & 27.05          & 35.71          & 23.69          & 25.90          & 34.73          & 26.06          & 28.15          \\
LLAMA2-CHAT                 & 7B                    & g                         & 36.76          & 35.02          & 35.41          & 38.36          & 34.87          & 35.63          & 28.57          & 36.97          & 35.27          & 28.57          & 32.93          & 32.13          & 33.89          & 34.62          & 34.58          \\
InternLM*                   & 7B                    & g                         & -              & -              & 49.00          & -              & -              & 49.20          & -              & -              & 40.50          & -              & -              & 49.40          & -              & -              & 47.10          \\
TigerBot-chat-v3            & 7B                    & g                         & {\ul 39.71}    & 45.99          & 44.59          & 32.88          & 52.49          & 48.20          & 14.29          & 41.82          & 36.23          & {\ul 39.29}    & 51.00          & 48.85          & 33.05          & 48.49          & 45.26          \\
Baichuan2-Chat              & 13B                   & g                         & 25.00          & 60.34          & 52.46          & 36.99          & 66.67          & 60.18          & 33.33          & \textbf{61.82} & \textbf{56.04} & {\ul 39.29}    & {\ul 65.06}    & {\ul 60.33}    & 33.47          & 63.81          & 57.43          \\
Ziya-LLaMA-v1*              & 13B                   & g                         & -              & -              & 43.30          & -              & -              & 36.90          & -              & -              & 34.30          & -              & -              & 41.20          & -              & -              & 39.30          \\
Qwen                        & 7B                    & g                         & 32.35          & \textbf{64.98} & \textbf{57.70} & {\ul 46.58}    & {\ul 68.20}    & {\ul 63.47}    & 23.81          & 55.76          & 49.28          & {\ul 39.29}    & 63.45          & 59.02          & {\ul 36.82}    & {\ul 64.54}    & {\ul 58.21}    \\ \midrule
FinGPT-v3 & 6B & f & 17.65 & 29.11 & 26.56 & 35.62 & 38.31 & 37.72 & 33.33 & 26.06 & 27.54 & 21.43 & 33.73 & 31.48 & 26.78 & 32.46 & 31.28 \\
ChatGLM2-AFAC2023Generation & 6B                    & f                         & 33.82          & 58.23          & 52.79          & 34.25          & 57.85          & 52.69          & {\ul 38.10}    & 44.24          & 43.00          & 37.50          & 56.22          & 52.79          & 35.56          & 54.76          & 51.00          \\
ChatGLM2-Yayi               & 6B                    & f                         & 36.76          & 54.01          & 50.16          & 39.73          & 56.70          & 52.99          & \textbf{40.48} & 40.00          & 40.10          & 32.14          & 53.82          & 49.84          & {\ul 36.82}    & 55.37          & 49.09          \\
Fin-Alpaca-LoRA-Linly            & 7B                    & f                         & 19.12          & 29.11          & 26.89          & 23.29          & 27.59          & 26.65          & 19.05          & 29.09          & 27.05          & 19.64          & 28.92          & 27.21          & 20.50          & 28.71          & 26.93          \\
DISC-FinLLM                 & 13B                   & f                         & 33.82          & 49.79          & 46.23          & 32.88          & 54.02          & 49.40          & 26.19          & 44.85          & 41.06          & 33.93          & 55.42          & 51.48          & 32.22          & 51.27          & 47.61          \\ 
Qwen-Yayi                   & 7B                    & f                         & 33.82          & 53.16          & 48.85          & 36.99          & 62.45          & 56.89          & 23.81          & 51.52          & 45.89          & 35.71          & 61.04          & 56.39          & 33.47          & 58.14          & 52.65          \\ \midrule
 NumLLM &7B& f & 32.06 & {\ul 63.80} & {\ul 56.72} & \textbf{47.67} & \textbf{68.97} & \textbf{64.31} & 26.19 & {\ul 57.94} & {\ul 51.50} & \textbf{44.29} & \textbf{65.38} & \textbf{61.51} & \textbf{38.74} & \textbf{65.40} & \textbf{59.25} \\
&  &  & (2.16) & (0.56) & (0.55) & (1.60) & (0.24) & (0.40) & (0.00) & (1.19) & (0.95) & (1.34) & (0.82) & (0.74) & (1.11) & (0.73) & (0.36)\\

\bottomrule

\end{tabular}
}
\label{table-FinEval上的总结果}
\end{table*}

\subsubsection{Implementation Details}
For hyperparameters mentioned in Section \ref{section:Fin-Textbooks}, we set $n_{min}=3, n_{max}=8, r_{\text{ins}}=0.05, r_{\text{NV}}=0.3 $. The experiments on hyperparameters can be found in the Appendix. In continual pre-training, we set the learning rate to be $5\times 10^{-5}$ and adjust it with the cosine annealing schedule during training. We set the block size to be 512 where the block size denotes the maximum length of the input sequence. We run the continual pre-training on 8 Tesla-V100-32G GPUs. The batch size per GPU is set to be 8. The number of total optimization steps is 6004 and the patience of early stopping is 5 epochs. For NumCT, we set the learning rate to be $5\times 10^{-5}$ and adjust it with the cosine annealing schedule during training.

\subsubsection{Baselines}

The baselines can be mainly categorized into two classes. The first class includes general-purpose LLMs that are able to answer financial questions. The second class includes financial LLMs that are fine-tuned from open-sourced LLMs on financial corpora.

The general-purpose LLMs for comparison include ChatGLM2-6B~\citep{GLM}, ChatGLM3-6B~\citep{GLM}, LLaMA-7B~\citep{LLaMA}, LLAMA2-7B-CHAT~\citep{LLAMA2}, Qwen-7B~\citep{Qwen}, InternLM-7B~\footnote{\url{https://github.com/InternLM/InternLM-techreport}}, Tigerbot-7B-chat-v3~\citep{TigerBot}, Baichuan2-13B-Chat~\citep{Baichuan2} and Ziya-LLaMA-13B-v1~\citep{Ziya}.

The financial LLMs for comparison include FinGPT-v3-6B~\citep{FinGPT}, ChatGLM2-6B-AFAC2023Generation, ChatGLM2-6B-Yayi, Qwen-7B-Yayi, Fin-Alpaca-LoRA-7B-Linly~\citep{Cornucopia} and DISC-FinLLM-13B~\citep{DISC-FinLLM}. ChatGLM2-6B-AFAC2023Generation is fine-tuned from ChatGLM2-6B with the instruction dataset AFAC2023Generation derived from the AFAC2023 competition in generation of financial market viewpoints~\footnote{\url{https://tianchi.aliyun.com/competition/entrance/532091/information}}. ChatGLM2-6B-Yayi is fine-tuned from ChatGLM2-6B with the instruction dataset constructed in Yayi. Qwen-7B-Yayi is fine-tuned from ChatGLM2-6B with the instruction dataset constructed in Yayi. DISC-FinLLM-13B refers to DISC-FinLLM-13B (consulting) which performs the best among the four variants proposed in the original work.

\subsection{Results on Financial Question Answering}

Experiment results are presented in Table~\ref{table-FinEval上的总结果}. The results of NumLLM are averaged over five independent NumCT runs. From Table~\ref{table-FinEval上的总结果}, we can find the following phenomena. 

Firstly, NumLLM outperforms all baselines in terms of overall accuracy, overall accuracy on numeric questions and overall accuracy on non-numeric questions.

Secondly, on numeric questions of the sub-domains, \mbox{NumLLM} outperforms Qwen on Finance, Economy and Certificate by a large margin. More specifically, NumLLM achieves accuracy gains of 5.00\%, 2.38\% and 1.09\%, respectively. Meanwhile, NumLLM is on par with Qwen on Accounting.

Thirdly, on non-numeric questions of the sub-domains,
NumLLM also outperforms Qwen on Finance, Economy and Certificate by a large margin. More specifically, \mbox{NumLLM} achieves accuracy gains of 1.97\%, 2.18\% and 0.77\%, respectively. Meanwhile, \mbox{NumLLM} keeps the second-best accuracy among all the compared models.
Please note that Qwen-Yayi is fine-tuned from the same foundation model as NumLLM but on different corpora. However, Qwen-Yayi achieves much lower scores than NumLLM.

\begin{figure}[t]
    \centering
    \includegraphics[width=0.5\linewidth]{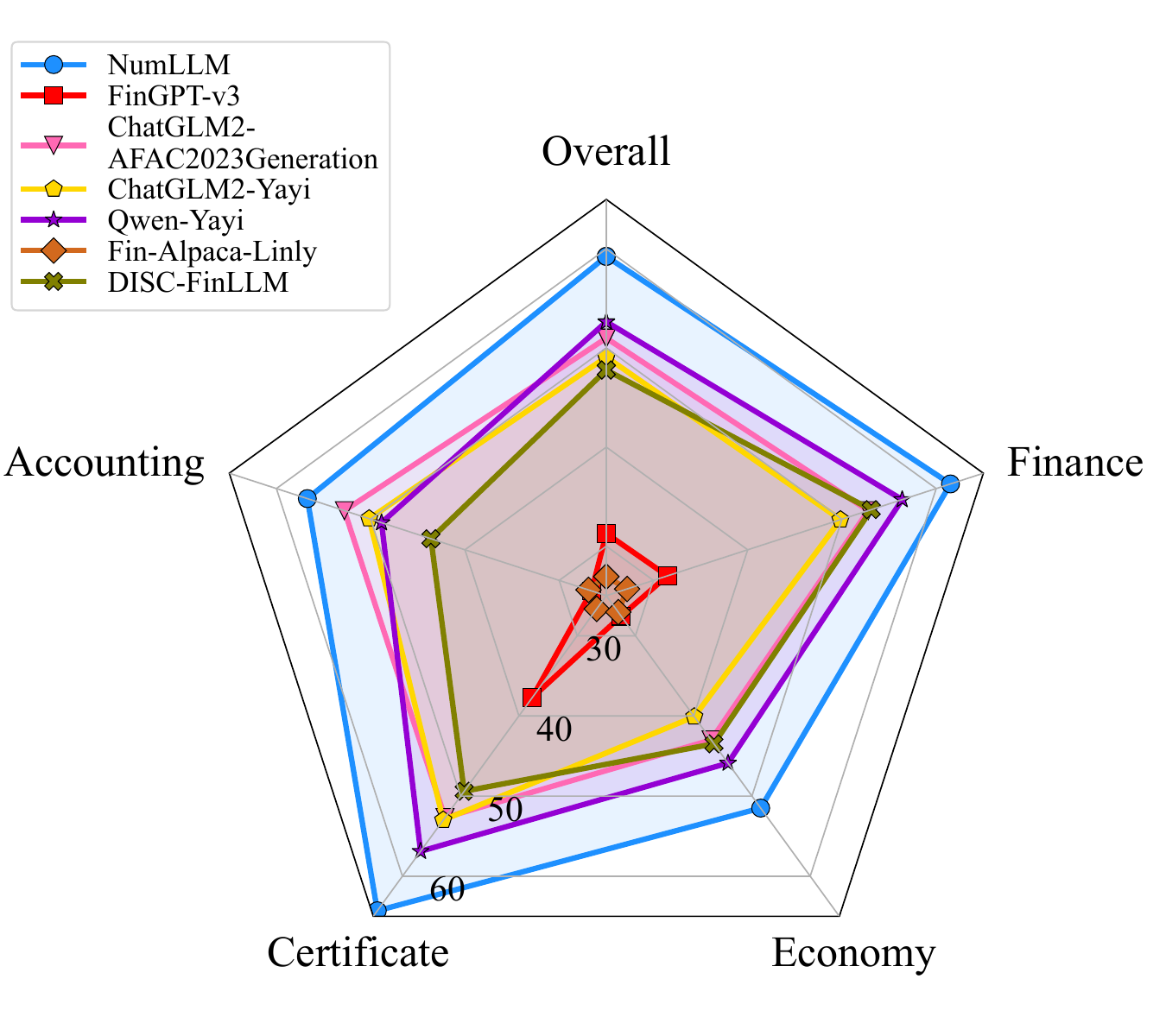}
    \caption{A radar graph for average (over numeric and non-numeric questions) accuracy (\%) of all financial LLMs  in all sub-domains.}
    \label{figure-radar}
\end{figure}

Finally, among all FinLLMs, NumLLM achieves the highest average accuracy in terms of overall results and results within each sub-domain. We can also observe this phenomenon from the radar graph in Figure~\ref{figure-radar}.

\begin{table}[!t]
\small
\centering 
\caption{Ablation study. Accuracy (\%) on FinEval overall. The numbers within parentheses are the standard deviations.}

\begin{tabular}{@{}l|rrr@{}}
\toprule
\multirow{2}{*}{model} & \multicolumn{3}{c}{Overall}                                 \\ 
    & \multicolumn{1}{c}{n}              & \multicolumn{1}{c}{non-n}          & \multicolumn{1}{c}{avg}   \\ \midrule
NumLLM (w/o NumCT)                          & \multicolumn{1}{l}{37.66}          & \multicolumn{1}{l}{64.90}          & \multicolumn{1}{l}{58.73} \\
NumLLM (w/o numeric choices)               & 37.24 (1.57)          & 63.85 (0.52)          & 58.33 (0.67)                     \\
NumLLM (w/o CP)                            & 31.52 (2.56)          & 63.85 (0.54)          & 57.14 (0.23)                     \\
NumLLM (sum-based mix)                               & 33.47 (0.59)          & 63.34 (0.61)          & 57.14 (0.57)                     \\
NumLLM (mean-based mix)                              & 38.49 (0.34)          & 64.07 (0.31)          & 58.76 (0.22)                     \\
NumLLM                                     & \textbf{38.74} (1.11) & \textbf{65.40} (0.73) & \textbf{59.25} (0.36)            \\ \bottomrule
\end{tabular}
\label{table-ablation}
\end{table}

\subsection{Ablation Study}
To study the effectiveness of each procedure during the construction of NumLLM, we conduct the ablation study by substituting each procedure with its variants or removing the procedure. The results are presented in Table~\ref{table-ablation}.

\subsubsection{Effectiveness of NumCT}
To verify the effectiveness of NumCT, we remove the LoRA module obtained by NumCT. Therefore, the foundation model is merged only with the LoRA module obtained by continual pre-training. The model obtained under this setting is denoted by NumLLM (w/o NumCT) in Table~\ref{table-ablation}. We can find that the accuracy of NumLLM (w/o NumCT) are 1.08\%, 0.50\% and 0.52\% lower than NumLLM on numeric questions, non-numeric questions and their average, respectively.

Moreover, we verify the effectiveness of numeric-masked choice generation within the procedure of NumCT. More specifically, we remove the step of numeric-masked choice generation when constructing NumCT instructions. For the target numeric variable in each instance, we transform the instance into a question by masking the numeric variable with a blank underline of four token length. The instruction is constructed by concatenating the necessary prompt constituents and the masked instance. The output is set to be the corresponding true value of the target numeric variable. The model obtained under this setting is denoted by NumLLM~(w/o numeric choices) in Table~\ref{table-ablation}. 
We can find that the accuracy of NumLLM~(w/o numeric choices) decreases by 1.50\%, 1.55\%, 0.92\% compared to that of NumLLM on numeric questions, non-numeric questions and their average, respectively. Similarly, the accuracy of NumLLM~(w/o numeric choices) decreases by 0.44\%, 1.05\%, 0.40\% compared to that of NumLLM~(w/o NumCT) on numeric questions, non-numeric questions and their average, respectively.

\subsubsection{Effectiveness of Continual Pre-Training}
To verify the necessity of conducting continual pre-training, we train a model which only performs NumCT with LoRA on Qwen but without LoRA for continual pre-training. 
The model obtained under this setting is denoted by \mbox{NumLLM~(w/o CP)} in Table~\ref{table-ablation}. We can find that the accuracy of \mbox{NumLLM~(w/o CP)} decreases by 7.22\%, 1.55\% and 2.11\% on numeric questions, non-numeric questions and their average, respectively.

\subsubsection{Effectiveness of SVD-based Method to Mix LoRA Modules}
To verify the effectiveness of the SVD-based method for mixing the two LoRA modules, we construct two variants of \mbox{NumLLM} for comparison. Specifically, we construct one variant using mean-based method for mixing LoRA modules, which adopts the $\Delta W_{\text{mean}}$ in Section \ref{section:mixing lora} as the full matrix of the mixed LoRA module. We construct the other variant using sum-based method for mixing LoRA modules, which adopts $\Delta W_{\text{sum}}=\Delta W_{\text{NumCT}}+\Delta W_{\text{CP}}$ as the full matrix of the mixed LoRA module.
These two variants are denoted by NumLLM (mean-based mix) and NumLLM (sum-based mix), respectively. From Table \ref{table-ablation},
we can find that \mbox{NumLLM (sum-based mix)} achieves the lowest accuracy among all three different mixing methods.
Furthermore, when compared to \mbox{NumLLM} (mean-based mix), \mbox{NumLLM} improves the accuracy by 0.25\%, 1.33\% and 0.48\% on numeric questions, non-numeric questions and the average result, respectively.
This proves the superiority of SVD-based method for mixing LoRA modules over mean-based method. One possible reason to explain this result is that because the ranks and training objectives are both different between continual pre-training and NumCT, the subspaces of $\Delta W_{\text{NumCT}}$ and $\Delta W_{\text{CP}}$ have different meanings which will result in noises in computing $\Delta W_{\text{mean}}$. But NumCT can mitigate the resulting noise through SVD, since SVD is an effective way for denoising~\citep{DBLP:journals/tcsv/GuoZZL16}.

\section{Conclusion}
In this paper, we propose a novel LLM, called \underline{num}eric-sensitive \underline{l}arge \underline{l}anguage \underline{m}odel~(NumLLM), for Chinese finance, which addresses the shortcoming of existing FinLLMs in understanding financial text when numeric variables are involved in questions.
Experiments on financial question-answering benchmark show that \mbox{NumLLM} can outperform existing FinLLMs to achieve the best performance. Applying our method for finance in other languages will be pursued in our future work.

\bibliography{references}
\end{document}